# A Sensitive Attribute based Clustering Method for k-anonymization

Pawan R Bhaladhare
Dept. of Information Technology
SNJB's College of Engineering, Chandwad, Dist. Nashik
E-mail: pawan_bh1@yahoo.com

Devesh C Jinwala
Dept. of Computer Engineering
SV National Institute of Technology, Surat, India
E-mail: dcjinwala@gmail.com

*Abstract—*
*In medical organizations large amount of personal data are collected and analyzed by the data miner or researcher, for further perusal. However, the data collected may contain sensitive information such as specific disease of a patient and should be kept confidential. Hence, the analysis of such data must ensure due checks that ensure protection against threats to the individual privacy. In this context, greater emphasis has now been given to the privacy preservation algorithms in data mining research. One of the approaches is anonymization approach that is able to protect private information; however, valuable information can be lost. Therefore, the main challenge is how to minimize the information loss during an anonymization process. The proposed method is grouping similar data together based on sensitive attribute and then anonymizes them. Our experimental results show the proposed method offers better outcomes with respect to information loss and execution time.*

## I. INTRODUCTION

With the upsurge in the proliferation of the information Technology in all walks of life and the subsequent exponential growth of data handled, it is not only becoming increasingly essential to analyze the same to derive useful information from therein, but also to ensure its justified and secure access and usage to/by individuals. Thus, on one side thread of research focuses on the efficient means of mining the data to gain useful knowledge, a parallel growing concern is to devise appropriate techniques for ensuring the privacy of the data even while allowing it to be used for mining.

For example, organizations such as hospitals contain medical database and they provide this database to the data miner or researcher for the purpose of analysis and research. The database consists of attributes such as name, age, gender, zip code and disease. The data miner will analyze the medical database to gain useful global health statistics. However, in the process the data miner or adversary may able to obtain sensitive information and in combination with an external database may try to obtain further personal attributes of an individual. This will lead to the threats to the disclosure of personal information of an individual. To avoid this many techniques have evolved but none of these provide a satisfactory solution.

Several latest research papers focuses on a range of data mining techniques such as non-cryptographic technique and cryptographic technique. The non-cryptographic technique contains inference problem, statistical database, k-anonymity, cluster analysis, classification, generalization, and association rule. The cryptographic techniques support secure multi-party computation. The aim of these techniques is to endorse the privacy. The problem with the non-cryptographic method results in a loss of information. On the other hand, cryptographic method provides accurate result but involve high computation and communication cost.

Non-cryptographic method are based on perturbation [23],[24],[25], k-anonymity[9][26][14][27][10][16][1], *l*-diversity[18] and t-closeness model[28], (α, k) anonymity[29]. The perturbation method is performed by adding noise [23], condensation [24] and swapping [25].The limitation of the perturbation method is that it corrupts the truthfulness of the released data. While in k-Anonymity the released data remain true. K-anonymized data is suffered from inference attack. To avoid inference attack on anonymized data, some anonymization techniques such as *l*-diversity and t-closeness have been proposed. The first algorithm for k-anonymity was proposed in saramati [1]. This method uses generalization on the quasi-identifier to build the anonymize table. Sweeney [30] uses the techniques such as generalization and suppression. In generalization, the attribute values are generalized for example, quasi-attribute age 33 could be generalized to a range such [31-40]. On the other hand, in suppression the quasi-attribute age 33 is replaced with 3*. This method reduces the risk of identification from the original database and or external available database. Variations to k-anonymity are top-down specialization and bottom-up generalization [31] [32]. A condensation based approach [33] has been used for the classification which from a cluster of records. A detail discussion about the personalized privacy is discussed in [34]. *l*-diversity method [18] was proposed which works on the sensitive attributes. A method have been discussed in detail in [19]

[35] for constructing the table. t-closeness model [28] is an enrichment to the concept of *l*-diversity. This method is also based on the sensitive attribute. The Earth Mover distance metric is used in order to enumerate the distance between the two distributions. Moreover, the *t*-closeness approach tends to be more effective than other privacy-preserving data mining methods for the numeric attributes. A condensation approach is discussed in [24] which generate the cluster to preserve the k-anonymity. This method is suffered from a large amount of information loss due to clustering of records.

Cryptographic method is applicable for the distributed database. In this method the data may be partitioned into horizontal database or vertical database. Due to which the records are extend across multiple entities. A broad overview on the cryptography is presented in [36]. A framework is illustrated in [37] for the secure multi-party computation problems. The secure multi-party computation methods contain methods such as secure sum, secure set union, intersection and scalar product [38]. Other methods includes Naïve Bayes Classifier [39], SVM Classifier [40]. Association rule mining [41], clustering [42] have been used in the horizontal partitioned data sets. While for vertical partitioned data methods such as Decision trees [43], SVM Classification [44], and *k*-means clustering [2] have been discussed.

One of the important methods for preserving the privacy is anonymity. The k-anonymity model proposed by Samarati [1] is a privacy preserving approach to protect the data. In this method, each record of a table is identical to at least (k-1) other record. Anonymity protects the data by hiding the details of the individual involved. Identity privacy and attribute privacy are the two parts which can protect the individual. The anonymization process uses generalization and suppression to ensure the privacy of the data. Generalization replaces a value with less specific but semantically consistent value. While suppression hides the data or does not release the entire value. Generalization is better in information extraction than suppression, which reduces the quality of the data and information loss. Consequently, the main difficulty of anonymization process is to maintain the data utility and privacy. Hiding the data reduces the data utility while disclosing the data reduces the privacy.

There is a tradeoff between the privacy and information loss. Therefore, a new anonymization approach is necessary to devise. To overcome the above issue, a sensitive attribute based clustering is proposed. This method sorts out all of the records in the table and then a cluster of records based on similar sensitive attribute is grouped. Many methods for clustering has been proposed in the literature [3][4][5][6][7]. The proposed method differs from Md. Enamul kabir et al. [8] in terms of information loss and the execution time and also with Byun et al. [3], Loukides et al. [7]. The proposed method diverges from the previous k-anonymization based clustering method in two different ways. First, our method make all cluster simultaneously like Md. Enamul kabir [8], while Byun et al. [3], Loukides et al. [7] create one cluster at a time. Second our proposed method takes less time than Md Enamul kabir. The performance of the proposed method is compared with the method proposed by Md. Enamul Kabir et al.[8].

The paper is divided into five sections. Section 2 illustrates the related work. Section 3 discusses the proposed approach. Section 4 illustrates the analysis of the algorithm. Section 5 shows the performance analysis. Finally a concluding remark and a future scope are presented.

## II.  RELATED WORK

The problem with preserving the privacy of an individual when data mining has gained much importance in recent years and due to this many algorithms have been proposed [9][10][12][13][14][15][16]. There are various issue involved with these algorithm such as high information loss [9][14][10][16] from predefined generalization hierarchies, measuring the total number of suppressions [17], the height of the generalization hierarchies [3][1] , size of the anonymized group [9][10] and information loss through anonymization[18]. Therefore, such metrics fail to detain security. Other works are in process to [19][20] enhance the protection by enforcing anonymized groups.

There are many clustering techniques that are used in the literature [3][4][6][7][10] which are used to protect the privacy of sensitive attributes. Byun [3] proposed the greedy k-member clustering algorithm. This algorithm builds the cluster by randomly selecting a record. Subsequently add the record in the cluster such that records have least information loss within the cluster. However, this algorithm is slow and sensitive to outliers. The experimental result illustrates that k-member algorithm causes less information loss than "Mondrian" anonymization technique [10]. Another clustering technique for k-anonymization has been proposed by Loukides and Shao [7]. This algorithm also chooses the seed of each cluster randomly. A user defined threshold is given to each cluster while adding the records in each cluster. Due to user defined threshold, this algorithm is less sensitive to outliers. The entire cluster is deleted, when the number of record is less than k. But the problem with this algorithm is that it is difficult to decide the user defined threshold. Also this algorithm may delete many records which may cause a high amount of information loss. Another algorithm for k-anonymization proposed by Chiu and Tsai[4] that adapts c-means clustering. This algorithm adds the records in the closest cluster based on weighted feature. If the cluster contain fewer than k records then those cluster should be merged with other large cluster. The only limitation of this algorithm is that it can only be used for quantitive quasi-identifier. Lin and Wei [6] proposed a one pass k-means clustering algorithm. The performance of this algorithm is better than Byun et al [3] with respect to information loss and execution time. This algorithm finds the closest cluster and assigns the records. If some cluster contains more the k records, then it removes the excess records and adds record to the cluster whose size is less than k. The drawback of this algorithm is that it has less execution time. Gonzalez [5] proposed the k-center clustering problem which finds k-clusters and minimizes the inter-cluster distance.

## III THE PROPOSED APPROACH

Clustering partitioned a set of records into groups such that records in the one group are more similar to the records of other groups. In this section, we present our new clustering algorithm that minimizes the information loss with respect to k-anonymity requirement. In the k-anonymity, the number of records in each equivalence class should be at least k and there is no restriction about the number of clusters. Therefore, the main objective of the clustering problem is to find the cluster which contains similar records and minimizes the information loss.

The central outline of the proposed algorithm is as follows.
We would like to anonymize the patient database based on sensitive attribute. Let us assume that we have attributes such as age, gender, Zip code and Occupation. The attribute such as age, gender and Zip code are consider as quasi-identifier, while the occupation attribute is consider as sensitive. Our objective is to create a cluster based on sensitive attribute. First we will sort all the records of the age attribute. Based on the sensitive attribute, a cluster is created. Next, we will find the minimum and maximum age in the cluster for a first sensitive attribute. Replace the minimum and maximum value with age attribute for all the records of the selected sensitive attribute. The detailed theoretical calculations are discussed in section 4.1. Moreover, the clusters are created such that the size of each cluster is greater than or equal to k and total information loss is at minimum. However, an experiment has been conducted to check the efficiency of the algorithm.

## IV ANALYSIS

### 4.1 Theoretical analysis

The main intention of the anonymization is used to measure the amount of information loss. There are many methods for calculating the information loss [3][9][21][6][22]. In this paper, the calculation of information loss is based on the Byun et al.[3].

Let us assume that a table consists of set of records with r quasi-identifier and s categorical attribute. Consider a cluster $\tau$ in $\mu$ consists of numerical and categorical attributes. Let $Nimax$ and $Nimin$ be the maximum and minimum values of the records in a cluster $\tau$ and $\mu Nimax$, $\mu Nimin$ be the maximum and minimum values of the records in $\mu$ with respect to the numerical attribute. While $Ucj$ be the union set of values in cluster $\tau$ with respect to the categorical attributes. $H(\xi cj)$ is the height of the taxonomy tree $\xi$. Therefore, the amount of information loss $IL(\psi)$ will be as

$$IL(\psi) = |\tau| \cdot \left( \sum_{i=1}^{r} ((Nimax-Nimin)/(\mu Nimax-\mu Nimin)) + \sum_{j=1}^{s} \frac{H(Ucj)}{H(\xi cj)} \right)$$

**Table 1 Patient table**

| Name | Age | Gender | Zip code | Disease |
|------|-----|--------|----------|---------|
| Ajay | 25 | Male | 443350 | Diabetes |
| Vijay | 26 | Male | 443351 | Cancer |
| Kamal | 27 | Male | 443352 | Flu |
| Rajesh | 36 | Male | 443350 | Hepatitis |
| Anjana | 40 | Female | 443350 | Hepatitis |
| Rajani | 39 | Female | 443350 | Hepatitis |

**Table 2 Anonymized table**

| Age | Gender | Zipcode | Disease | Count |
|-----|--------|---------|---------|-------|
| 25 | Person | 4433** | Diabetes | 1 |

| 26 | Person | 4433** | Cancer | 1 |
| 27 | Person | 4433** | Flu | 1 |
| 36-40 | Person | 4433** | Hepatitis | 3 |

For example consider the Table 1 the patient table and the anonymized Table 2. The anonymized table consists of four clusters. Attribute such as Age, Gender, Zip code and Disease, where Age and Zip code are quantitative attribute while the Disease is a categorical attribute. Also consider the height of the taxonomy tree for the Zip code, Gender is one level only. In the first second and third cluster the maximum and the minimum values are 25, 26, and 27 respectively. While in the fourth cluster these values are 40 and 36. Also, the maximum and minimum value for the whole table is 40 and 25 respectively. Then the total information loss for the anonymized table will be 17.28 by using the formula given by Byun et al.[3]. However, the information loss as per the method suggested by Md. Enamul K. et al.[8] is 17.70. These shows that the proposed method has less information loss than the method describe by Md. Enamul K. et al.[8]. The main intention of clustering techniques is to construct the cluster in such a way that the total information loss will be less.

### 4.2 Time complexity

This algorithm sort all records based on the quasi-identifier and create partition all records into n/k group. The time its takes to sort all the records is O(n logn). While the time for clustering is $O(\frac{n^2}{k})$. Where n is the total number of records in the dataset. The clusters are created based on the sensitive attribute. The cluster contains at least k records. This process of creating a cluster takes minimum information loss.

### V PERFORMANCE ANALYSIS

### 5.1 Test application

The basic idea of the proposed algorithm is as follows.

Input:- Medical database which consist of Table and having the attribute such as Name, Age, Gender, Zip code and Occupation as a sensitive attribute.
Output:- Anonymized Table
1) Load the Database
2) Remove the identifier from the table
3) Create a new Database
4) Copy the contents of old Database to new.
5) Copy the Occupation into an array(Occupation[ ]).
6) While not EOF // sort according to Occupation and find minimum and maximum age in a cluster.
7) {
    i=0
    While i< sizeof(Occupation[ ])
        If minage[i]>rs.age
            Minage[i]=rs.age
        If maxAge[i]<rs.age
            maxAge[i]=rs.age
        rs.gender="Person"
        cntCluster=cntCluster+1
        rs.age=minAge[i] & " - " & maxAge[i]
        rs1.zipcode = Left(rs.zipcode,   Len(rs.zipcode) - 2) & "**"
    wend
                                        }
8) While i < size  // to find information loss
        T1sum = cntcluster(i) * (((maxAge(i) - minAge(i)) / (Tmax - Tmin)) + 1 + 1 + (i + 1) / size)
    Tsum = Tsum + T1sum
    i = i + 1
    Wend

### 5.2 Metrics for evaluation

The objective of the experiment is to investigate information loss and execution time. Evaluating the proposed algorithm with Md. Enamul Kabir et al. algorithm [8], showed that the proposed method causes less information loss and execution time.

### 5.3 Experimental setup

The objective of the experiment is to investigate information loss and execution time. Evaluating the proposed algorithm with Md. Enamul Kabir et al. algorithm [8], showed that the proposed method causes less information loss and execution time.  The experiment is implemented in VB 6.0 and MS Access and run on 3.2 GHz Intel Core 2 Duo Processor machine with 2 GB RAM . The Microsoft Windows XP Professional is used as an operating system. An adult dataset from the UCI Machine Learning Repository [11] is used. There are 32561 records used in the experiment. An adult dataset is used as a standards dataset for checking the performance of the k-anonymity algorithm. We retain only attribute such as Age, Gender, Zip code and Occupation. Among these Age and zip code are numeric attribute whereas Gender and Occupation are the categorical attribute. We consider occupation as the sensitive attribute in the table.

### 5.4 Result

Initially, the experiment is conducted for the only six records of the original table for figure 1 and figure 2. The proposed method produce four cluster, while a systematic clustering construct two cluster. Figure 1 show the execution time for the systematic clustering and the proposed method. The result show that execution time for the proposed method is less than the systematic clustering method proposed by Md. Enamul Kabir et al.[8]. Figure 2 illustrates the information loss for the systematic clustering and the proposed method.  It seems that the proposed method have less information loss than the systematic clustering proposed by Md. Enamul Kabir et al.[8].

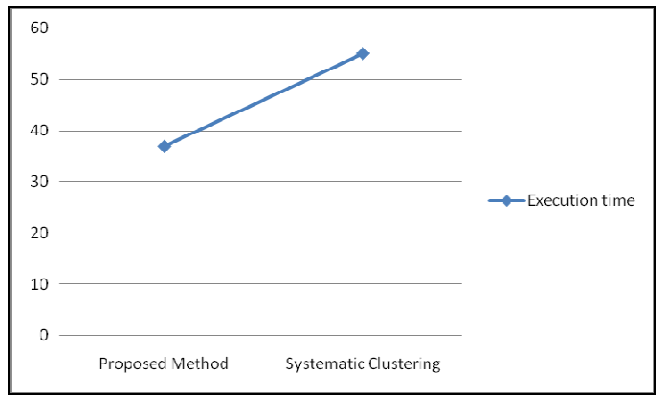

**Fig 1 Execution time for the Systematic clustering and the Proposed Method**

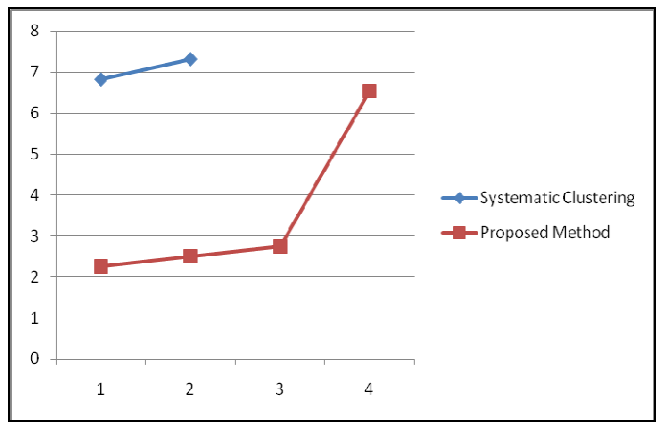

**Fig 2 Information loss for the Systematic clustering and the Proposed Method**

The experiments are conducted on four different scenarios with the information loss and the cluster sizes are collected for each run. Figure 3 illustrate the information loss for the proposed method. As the clustering is based on a sensitive attribute, the result shows that the cluster size and the information loss are directly proportional to each other.

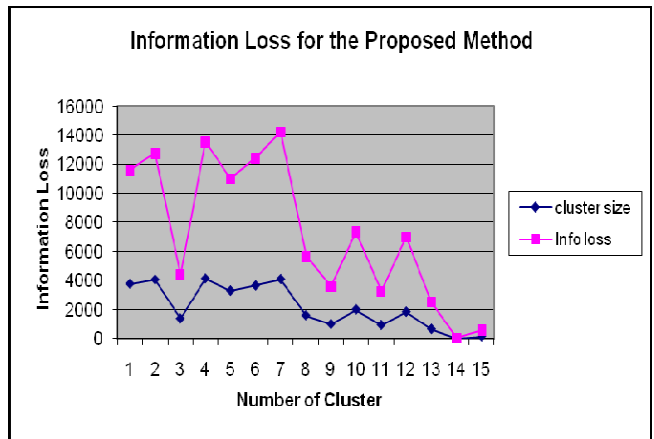

**Fig 3 Information loss for the Proposed Method**

Figure 4 shows the information loss for the Systematic clustering proposed by Md. Enamul K. et al[8].

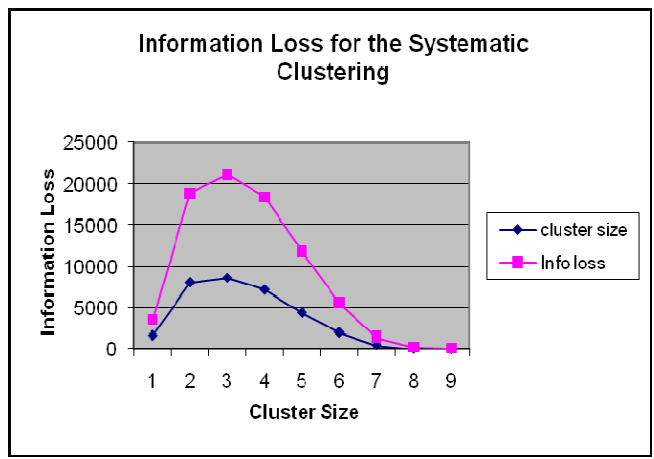

**Fig 4 Information loss for the Systematic clustering**

Figure 5 shows the comparison of information loss for systematic method and the proposed method. The result shows that the proposed method has less information loss than the systematic clustering [8].

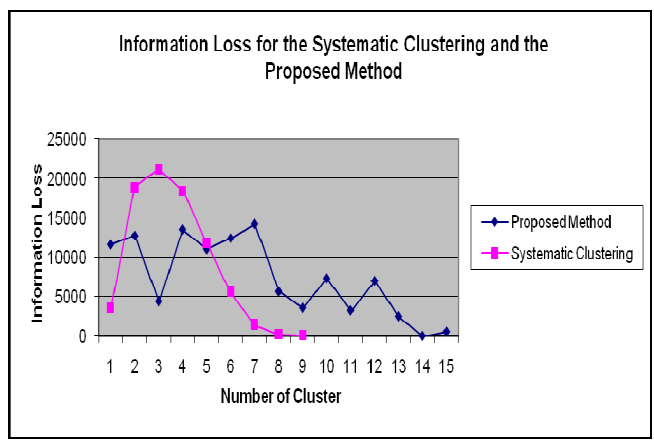

**Fig 5 Information loss for the Systematic clustering and the Proposed Method**

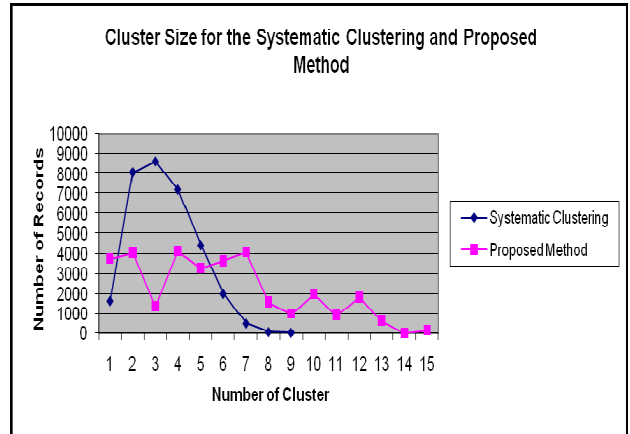

**Fig 6 Number of Cluster for the Systematic clustering and the Proposed Method**

Figure 6 show the comparison of the cluster size for the systematic clustering [8] and the proposed method. It shows that our clusters are of nearby of equal size. While the cluster size of systematic clustering proposed by Md.Enamul K. et al.[8] are of different size. With all different scenario we can conclude that the proposed method is better than the systematic clustering proposed by Md. Enamul K. et al.[8].

**VI CONCLUSION**

In contemporary society, privacy preserving data mining is useful in various applications worldwide in scenarios such as medical database analysis, market analysis and financial analysis. Mining these databases include private and sensitive information about an individual. The main intention is to minimize information loss and data utility, while also protecting the sensitive attributes and private information of an individual. In this paper, a sensitive attribute based clustering approach for k-anonymization was proposed, which shows comparable result with respect to information loss and execution time. Based on the investigations of the information loss and data utility with respect to the current privacy preserving approaches, the following probable research pointers were identified and are intended to be pursued. To investigate new non-cryptographic method: To protect the data many methods are modifying the quasi-identifier data and some are using rules to hide the data from disclosure. Currently, quasi-identifier has to be modified in order for no information to be lost and data utility to be maintained; as a result new methods need to be developed without any modification. To investigate a hybrid approach: Many algorithms are based on classification approaches and some are based on clustering approaches. While no work has been done on the combination of classification and clustering. Such hybrid approach can potentially provide new and better ways to protect the privacy.